\documentclass{article}

\usepackage[latin1]{inputenc}
\usepackage{graphicx}
\usepackage{amsmath}
\usepackage{amsfonts}
\usepackage{amssymb}
\usepackage{cite}

\title{Selective Response of Mesoporous Silicon to Adsorbants with Nitro Groups}

\author{John A. McLeod\footnote{Department of Physics and Engineering Physics, University of Saskatchewan, 116 Science Place, Saskatoon, Saskatchewan S7N 5E2, Canada}~\footnote{e-mail: \texttt{john.mcleod@usask.ca}} \and 
Ernst Z. Kurmaev\footnote{Institute of Metal Physics, Russian Academy of Sciences-Ural Division, 620990 Yekaterinburg, Russia} \and 
Peter V. Sushko\footnote{Department of Physics and Astronomy and the London Centre for Nanotechnology, University College London, Gower Street, London WC1E 6BT, United Kingdom, and WPI-Advanced Institute for Materials Research, Tohoku University, Sendai, Japan}~\footnote{e-mail: \texttt{p.sushko@ucl.ac.uk}} \and 
Teak D. Boyko\footnote{Department of Physics and Engineering Physics, University of Saskatchewan, 116 Science Place, Saskatoon, Saskatchewan S7N 5E2, Canada} \and 
Igor A. Levitsky\footnote{Emitech, Inc., 150 Harvard St., Fall River, Massachusetts 02720, and Department of Chemistry, University of Rhode Island, Kingston, Rhode Island 02881, USA} \and 
Alexander Moewes\footnote{Department of Physics and Engineering Physics, University of Saskatchewan, 116 Science Place, Saskatoon, Saskatchewan S7N 5E2, Canada}}

\begin{document}
\maketitle
\begin{abstract}
We demonstrate that the electronic structure of mesoporous silicon is affected by adsorption of nitro-based explosive molecules in a compound-selective manner. This selective response is demonstrated by probing the adsorption of two nitro-based molecular explosives (trinitrotoluene and cyclotrimethylenetrinitramine) and a nonexplosive nitro-based aromatic molecule (nitrotoluene) on mesoporous silicon using soft X-ray spectroscopy. The Si atoms strongly interact with adsorbed molecules to form Si-O and Si-N bonds, as evident from the large shifts in emission energy present in the Si \textit{L}$_{2,3}$ X-ray emission spectroscopy (XES) measurements. Furthermore, we find that the energy gap of mesoporous silicon changes depending on the adsorbant, as estimated from the Si \textit{L}$_{2,3}$ XES and $2p$ X-ray absorption spectroscopy (XAS) measurements. Our \textit{ab initio} molecular dynamics calculations of model compounds suggest that these changes are due to spontaneous breaking of the nitro groups upon contacting surface Si atoms. This compound-selective change in electronic structure may provide a powerful tool for the detection and identification of trace quantities of airborne explosive molecules.
\end{abstract}

\section{Introduction}

Several methods for detecting trace quantities of explosives have been developed in recent years.~\cite{yinon02} However, a simple, compact, sensitive, selective, non-invasive, and low-cost sensor for explosives is an ideal that has not yet been achieved. For example, a laser can be combined with a gas chromatographer in order to induce fragmentation of a target material and analyze its gas-phase molecular fragments,~\cite{steinfeld98} but this is not suited for non-destructive testing, or monitoring of suspected material \textit{in situ} (i.e. in an airport luggage scanner). Another common technique, pulsed neutron analysis, can be used to identify elements and quantify the bulk C/O and N/O ratios characteristic to explosives.~\cite{vourvopoulos01} This technique is fast and can easily be applied to bulk materials, however, pulsed neutron sources and $\gamma$-ray detectors are required.~\cite{vourvopoulos01} These devices can be made small enough to be deployed in field situations, but they are expensive and the detection procedure involves careful positioning of the device with respect to the suspect material. Hand held detectors can be realized by utilizing ion mobility spectrometry.  However, issues such as maintaining a stable ion source and avoiding false positives from gas-phase ion-molecule fragments are still to be resolved.~\cite{ewing01} Single-walled carbon nanotube field effect transistors have shown changes in resistivity in the presence of single molecules like glucose and DNA,~\cite{allen_07,cella_10} but their sensitivity to nitroexplosives has not yet been demonstrated. Recently, a solid acid catalyst has been shown to be sensitive to triacetone tetraoxide vapour,~\cite{lin_10} its sensitivity to other explosives has again not yet been demonstrated.

Optical spectroscopy appears to be a good candidate for detection of explosives, since both sources and detectors in this regime are typically not expensive and widely available. However many spectroscopic techniques (such as Raman and IR spectroscopy) require relatively large quantities of a material, positioned between the source and the detector, which creates difficulties in practical applications of these techniques. For example, the low vapour pressure of most explosives \cite{pushkarsky06} makes accurately detecting these compounds through adsorption or Raman spectroscopy quite difficult,~\cite{steinfeld98} since the signal-to-noise ratio in practical settings is often rather low. Surface-enhanced Raman spectroscopy was recently used to detect low concentrations of dinitrotoluene (DNT) and trinitrotoluene (TNT) adsorbed in alumina nanopores containing gold nanoparticles,~\cite{ko_09} although the detection sensitivity is highly dependent on the surface properties.~\cite{steinfeld98}

So far, the most sensitive method for detecting nitroexplosives is the luminescence quenching of emissive conjugated polymers (or amplifying polymers) developed by the Swager \textit{et al}.~\cite{yang98} This approach is capable of detecting TNT vapours in the particle per trillion (ppt) range and even lower. More recent research has found techniques for improving the fluorescence quenching of conjugated polymer layers and semiconducting organic polymers exposed to nitroexplosive vapour,~\cite{naddo_07,rose_05,yang_10,narayanan_08} and conjugated polymer sensitivity to a large number of nitroexplosives in solution has also been demonstrated.~\cite{sanchez_08} Nevertheless, the selectivity of this method is not sufficient as many non-explosive nitroaromatic compounds can also induce luminescence quenching.  Thus, there is clearly a need for a material that can quickly and selectively either accumulate molecules of the explosives at sufficiently high densities or noticeably change its properties under the influence of these molecules and ultimately detect and identify the specific compound present.

A promising method for tracing concealed explosives is detecting the adsorption and desorption of explosive molecules on surfaces of porous semiconductors.~\cite{levitsky07} Porous silicon (PSi) is an attractive material for the chemical sensing of vapours because it has high surface area and a variety of transduction mechanisms associated with its electrical, optical, and chemical properties. For example, it has been shown that the photoluminescence of PSi changes depending on chemical treatment due to surface effects and/or nanostructured silicon crystallites,~\cite{theiss_97} and the adsorbed molecules on the silicon surface.~\cite{chun_93,lauerhaas_92,song_97,krawiec_96,lauerhaas_93,lin_97} The resistivity of PSi can be modified by several orders of magnitude by the adsorption of dielectric liquids.~\cite{timoshenko01} Finally, PSi is readily functionalized by a large variety of molecular structures \cite{bjorklund_97,buriak_98,buriak_99,kim_97,kim_98} and is highly chemically reactive.~\cite{maley_10} It has also been suggested that nanoporous materials can preferentially adsorb different species of organic molecules,~\cite{dubbeldam_08} indicating that the right preparation of PSi could serve as a selective detector of explosive molecules.

PSi can adsorb large quantities of molecules on its surface  due to a high surface-to-volume ratio of up to 1000 m$^2$/cm$^3$ and activated diffusion of surface molecules into the pores,~\cite{valiullin_05} which can be exploited in order to control oxidation reaction and/or molecular intake rates.~\cite{mawhinney_97,jarvis_08} This, together with the availability of silicon-based microfabrication technologies, makes porous Si one of the most promising candidates for vapour sensing; indeed the possibility of using PSi in optical sensors for vapour detection has been intensively studied for the past decade. The adsorption of the target molecules into the silicon pores modifies the refractive index and consequently the optical properties of PSi. Optical sensors based on porous mono/double layers, Bragg mirrors, luminescent and reflective microcavities (MCs) have been reported in literature.~\cite{snow99,saarinen05,mulloni00,pacholski05,stefano04,kolobov10,levitsky07} Also, it should be noted that TNT detection with the use of a PSi microcavity infiltrated with a fluorescent sensory polymer has previously been studied.~\cite{levitsky07}
All the above features of PSi are critical for detection of trace levels of explosives, such as TNT and cyclotrimethylenetrinitramine (RDX), which exhibit very low pressures of saturated vapours (in the ppb to ppt range).~\cite{pushkarsky06}

Finally, PSi has previously been shown to have sub-ppm sensitivity to NO and NO$_2$ gas,~\cite{harper_96} and an NO$_2$ sensor using changes in conductance of PSi has been demonstrated to work well in a range of ambient temperatures and humidities.~\cite{boarino_00,bratto_01,pancheri_03}. It has also been demonstrated that aged PSi shows greater reversibility to NO$_2$ detection and less sensitivity to air moisture,~\cite{pancheri_04} suggesting that PSi may be a very practical field sensor. Finally, there is some chemisorption occuring during the interaction of NO$_2$ and PSi \cite{sharov_05} and the surface of the PSi changes upon nitridization and oxidization during this process. This suggests that molecules NO$_2$ groups (such as TNT or RDX, for example) will show some reactivity to PSi, and will be adsorbed on a PSi surface. It is therefore possible that the adsorbed molecule may functionalize the PSi surface and induce molecule-specific changes to the surface electronic structure.

\begin{figure}
\includegraphics[width=3in]{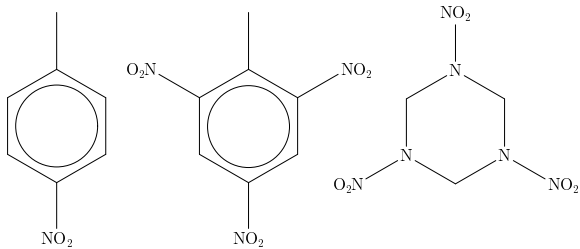}
\caption{\label{fig:molecules} A stylized schematic of (from left to right) para-nitrotoluene (NT), trinitrotoluene (TNT), and cyclotrimethylenetrinitramine (RDX).}
\end{figure}

To evaluate the potential of PSi for selective detection of explosives, it is important to study the details of the chemical interaction of silicon atoms with nitroaromatic and nitroamine molecules.  In this article we present soft X-ray spectroscopy measurements, a technique which offers direct, atom- and orbital-selective probe of local chemical bonding, to demonstrate how the local bonding characteristics of the Si atoms in the PSi structure are changed after exposure to nitrotoluene (NT), TNT, and RDX (shown in Figure \ref{fig:molecules}). Finally, X-ray spectroscopy measurements shed light on the shift of the band edges of PSi affected by these adsorbants. The experimental observations are corroborated by \textit{ab initio} molecular dynamics simulations and static energy minimization simulations, which reveal atomic-scale mechanisms of the adsorption and decomposition of model molecules at amorphous silicon surfaces and the corresponding modifications to the electronic structure of the combined system.

\section{Results and Discussion}

\begin{figure}
\includegraphics[width=3in]{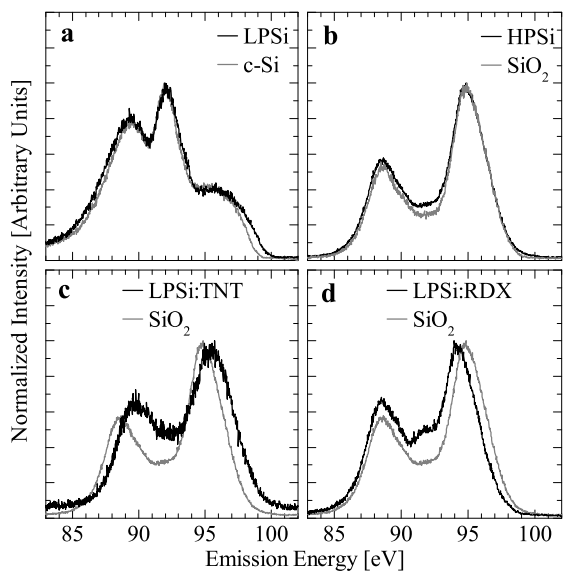}
\caption{\label{fig:silicon} The silicon \textit{L}$_{2,3}$ XES spectra. Panel \textbf{a} shows the similarity between LPSi and crystalline silicon (c-Si), suggesting a low amount of oxidation in LPSi. Panel \textbf{b} shows the similarity between HPSi and SiO$_2$, suggesting that HPSi is almost fully oxidized. Inset \textbf{c} shows how the spectrum of LPSi:TNT is shifted away from the spectrum of SiO$_2$. Panel \textbf{d} shows how the spectrum of LPSi:RDX is shifted in the opposite direction from the spectrum of LPSi:TNT, towards lower energies. (Note: A colour versions of this and other figures in this manuscript are available in Supporting Information.)}
\end{figure}

\paragraph{Silicon X-ray Spectroscopy.} It is well known that PSi as a chemically active, high surface-to-volume ratio material can be easily oxidized.~\cite{bisi00} The microporous (with a pore size greater than 50 nm), mesoporous (pore size of 5-50 nm) or nanoporous (pore size less than 5 nm) silicon layers can be bare, partially oxidized (when Si domains are embedded into an oxide matrix), or fully oxidized. These different forms of PSi have distinctly different physical properties, and their characteristics may vary during ageing due to oxidation and hydroxylation processes.~\cite{mizsei07} With this in mind, we examine the Si \textit{L}$_{2,3}$ XES spectra of LPSi and HPSi. As shown in Figure \ref{fig:silicon}, the Si \textit{L}$_{2,3}$ XES spectrum of LPSi is almost the same as that of bulk crystalline Si (c-Si), whereas the Si \textit{L}$_{2,3}$ XES spectrum of HPSi is the almost same as the spectrum of SiO$_2$. 
This means that HPSi is, for all practical purposes, fully oxidized and its surface is wholly composed of SiO$_2$. In fact, our spectroscopy measurements of NT, TNT, and RDX adsorbed on HPSi show the same trends as our corresponding measurements on LPSi, but the magnitude of the spectral changes is smaller (see Supporting Information), which we attribute to the lower reactivity of the HPSi surface. Hence we do not consider HPSi in the following discussion.

The Si \textit{L}$_{2,3}$ XES spectrum of LPSi shows that it is primarily pure silicon within 60 nm of the surface despite being oxidized for 20 minutes at 900 $^\circ$C. To rationalize this result, we note that the volume of SiO$_x$ ($0<x$) per Si atom is larger than that of pure silicon, which leads to build up of the lattice stress during Si oxidation and which is known to decrease the oxidation rate and result in formation of rough films.~\cite{Wang_1991} At temperatures above 1050 $^\circ$C, viscous flow of SiO$_2$ suppresses this stress and its effect on the oxidation rate.~\cite{Wang_1991} However, in our case, the viscous flow contribution is negligible and, consequently, accumulated stress suppresses further oxidation. Moreover, earlier experimental~\cite{Kao1} and theoretical~\cite{Kao2} studies have demonstrated that oxidation of concave surfaces, a typical geometry for pores in LPSi, progresses at a much lower rate than oxidation of flat surfaces. Thus, it is reasonable to suggest that the oxidized areas primarily form ``islands'' in the regions between pores, leaving the Si inside the pores exposed. Indeed, previous atomic force microscopy characterization of oxidized PSi shows that the oxide layer tends to form on the needle-like peaks on the PSi surface, rather than in the pores.~\cite{pap_05} An additional consideration is that the low pore density might allow an oxide layer to coat the entire surface, sealing off the tops of the pores.~\cite{barla_86} However, after the LPSi is cooled, the induced stress on this layer could fracture it, thus exposing the pure silicon pores.

A dramatic change is seen in the Si \textit{L}$_{2,3}$ XES spectra after exposing LPSi to nitro-based molecules. The fine structure in the spectrum of untreated LPSi is strongly modified and instead of the three features typical for c-Si, the spectra of LPSi exposed to the nitrogen compounds consist of two main peaks which are closer to the spectrum of SiO$_2$. Of the three treated materials, the spectrum of the substrate exposed to NT (LPSi:NT) is the most similar to that of SiO$_2$ (in energy position and peak ratio). The spectrum of the substrate exposed to TNT (LPSi:TNT) shows an energy shift compared to SiO$_2$, with the high energy XES peak shifted to higher energies (see Figure \ref{fig:silicon}). The spectrum of the substrate exposed to RDX (LPSi:RDX) is shifted to lower energies with respect to that of SiO$_2$, i.e. shows the opposite shift in energy compared to LPSi:TNT. Thus, we conclude that the surface of LPSi interacts strongly with the adsorbants and the effect of this interaction on the Si electronic structure depends on the molecule.

We would like to stress that the large magnitude of the energy shifts evident in the Si \textit{L}$_{2,3}$ XES spectra is very surprising, especially since the LPSi substrate was only exposed to a vapour of NT, TNT, or RDX molecules. These shifts are also seen somewhat in HPSi after exposure to these molecules (the shift is less apparent, due to the aforementioned lower chemical activity, but the trends are consistent with those seen in LPSi). The energy shifts in the Si \textit{L}$_{2,3}$ XES spectra make a very strong case for the sensitivity of the electronic properties of PSi to adsorbed molecules.

\begin{figure}
\includegraphics[width=3in]{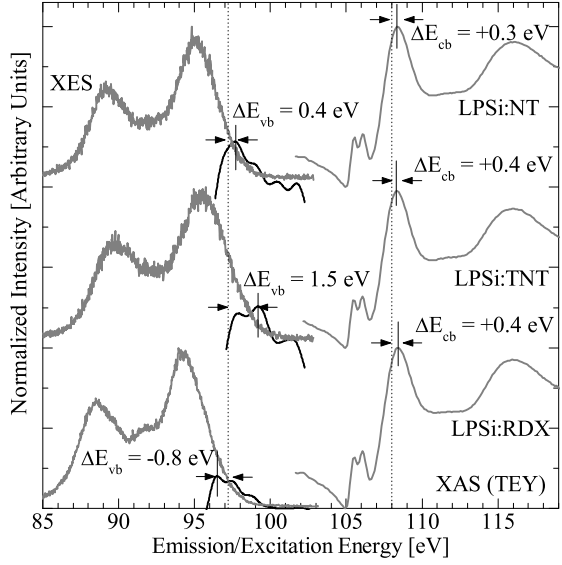}
\caption{\label{fig:silicongap} The silicon \textit{L}$_{2,3}$ XES and \textit{2p} XAS (TFY) spectra. The upper valence band edge of SiO$_2$ (estimated using the second derivative of the Si \textit{L}$_{2,3}$ XES spectrum of SiO$_2$) is shown by the dotted line at $\sim$ 97.7 eV, and the main peak of the Si \textit{2p} XAS spectrum of SiO$_2$ is shown by the dotted line at $\sim$ 108 eV. The pre-edge XAS spectral features at $\sim$ 106 eV are aligned in all XAS spectra, including that of SiO$_2$. The second derivative of the Si \textit{L}$_{2,3}$ XES spectra near the valence band edge is plotted on top of the appropriate Si \textit{L}$_{2,3}$ XES spectra for the treated LPSi samples.}
\end{figure}

\paragraph{Surface energy gaps.} We can estimate the changes in the surface energy gap by comparing the Si \textit{L}$_{2,3}$ XES and \textit{2p} XAS spectra, as shown in Figure \ref{fig:silicongap}. Note that while the 60 nm penetration depth of the Si \textit{L}$_{2,3}$ XES measurements would normally make this a bulk measurement, because the pores are usually a few hundred nm deep \cite{theiss_97,letant_04} and, more importantly, because the Si \textit{L}$_{2,3}$ XES clearly shows a sensitivity to the vapour treatment of the samples, we are justified in considering the Si \textit{L}$_{2,3}$ XES as a surface sensitive measurement. Using the main peaks of the second derivative of the complementary XES and XAS spectra of the oxygen \textit{K} edge has been shown to give a good estimate of the energy gap in the case of various metal oxides.~\cite{kurmaev08} 

Applying this method to the silicon \textit{L}$_{2,3}$ edge is problematic, since there are strong pre-edge features in the \textit{2p} XAS (around 107 to 109 eV), which are either due to the final state core hole \cite{ogasawara98} or exitons.~\cite{buczko00} To avoid this complication, we have used the main peak of the second derivative of the \textit{L}$_{2,3}$ XES spectrum as an estimate for the valence band edge, and have chosen the main peak of the \textit{2p} XAS as a reference for the lower conduction band edge --- i.e., we assume that the main peak of the Si \textit{2p} XAS is the same energy above the true conduction band edge for all LPSi samples. The energy separation between the peak in the second derivative of the Si \textit{L}$_{2,3}$ XES and the main peak of the Si \textit{2p} XAS features is 10.3 eV, while the energy gap of amorphous SiO$_2$ is closer to 8.9 eV.~\cite{distefano71} We stress that the peak in the Si \textit{2p} XAS does not correspond to the bottom of the conduction band, and, therefore, this method does not provide an estimate of the actual energy gap. However, comparing the shifts in this feature and in the peak in the second derivative of the Si \textit{L}$_{2,3}$ XES between the different materials should give a reasonably accurate estimate of the changes in the energy gap.

\begin{figure}
\includegraphics[width=3in]{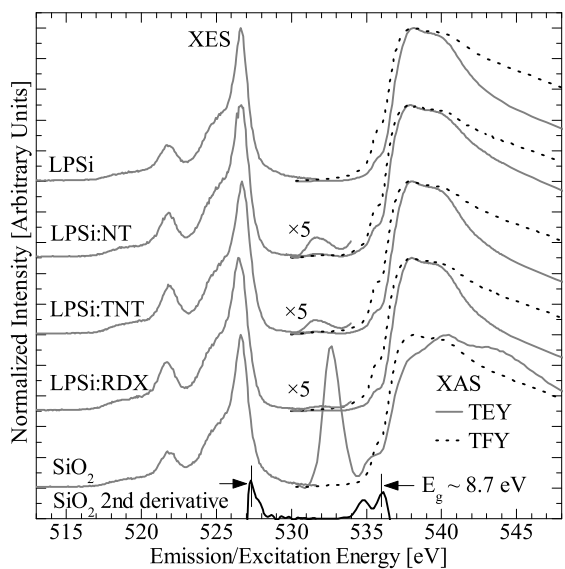}
\caption{\label{fig:oxygen} The oxygen \textit{K} XES and \textit{1s} XAS spectra for LPSi treated with various adsorbants compared to SiO$_2$. The energy gap of SiO$_2$ is estimated using the second derivative of the XES and XAS spectra. The pre-edge features in the XAS spectrum for the treated LPSi samples has been increased in intensity by a factor of 5. Both the bulk sensitive TFY and surface sensitive TEY modes are plotted for the O \textit{1s} XAS.}
\end{figure}

The second derivative of the Si \textit{L}$_{2,3}$ XES spectra near the valence band edge is shown in Figure \ref{fig:silicongap}, where it is superimposed on the measured Si \textit{L}$_{2,3}$ XES spectra. The valence band edge of LPSi:NT, as estimated from the peak in the second derivative of the Si \textit{L}$_{2,3}$ XES spectrum, is about 0.4 eV higher than that of SiO$_2$, while the bottom of the conduction band, as estimated from the peak in the Si \textit{2p} XAS spectrum, also appears at slightly higher energies (by about 0.3 eV). This suggests that the surface energy gap of LPSi:NT is roughly 0.1 eV smaller than that of SiO$_2$. In the case of LPSi:TNT, the valence band edge, referenced to the Si \textit{2p} core states, is 1.5 eV higher than in SiO$_2$, while the bottom of the conduction band is only 0.4 eV higher than that of SiO$_2$. Thus, the surface energy gap of LPSi:TNT is roughly 1.1 eV smaller than that of SiO$_2$. Similar analysis of the LPSi:RDX spectra suggests that the energy gap in this system appears to be 1.2 eV larger than that in SiO$_2$. These dramatic changes are significant, but not unexpected for this system. It is clear from Figure \ref{fig:silicon} and Figure \ref{fig:oxygen} that untreated LPSi has a energy gap close to that of c-Si in the bulk and that of SiO$_2$ on the surface, so there is a large change in electronic structure close to the surface. Indeed, changes in the porosity of pure silicon can affect the optical gap by $\sim$0.3 eV,~\cite{rotaru99} and the energy gap of various polymorphs of SiO$_2$ can vary by as much as 4 eV.~\cite{xu91} In light of this, since we are probing the interface of c-Si with SiO$_2$ and Si bonded to various organic functional groups (possibilities include nitro, amine, methyl, and aromatic groups), the large change in surface properties is not entirely unexpected.

The large energy shifts evident in the Si \textit{L}$_{2,3}$ XES of treated LPSi suggest that the Si surface strongly interacts with the adsorbed molecules. What is the nature of this interaction? Simply based on the structure of NT, TNT, and RDX we can suggest that it is driven by the formation of Si-O, Si-N, and/or Si-C bonds at the surface.

\begin{figure}
\includegraphics[width=3in]{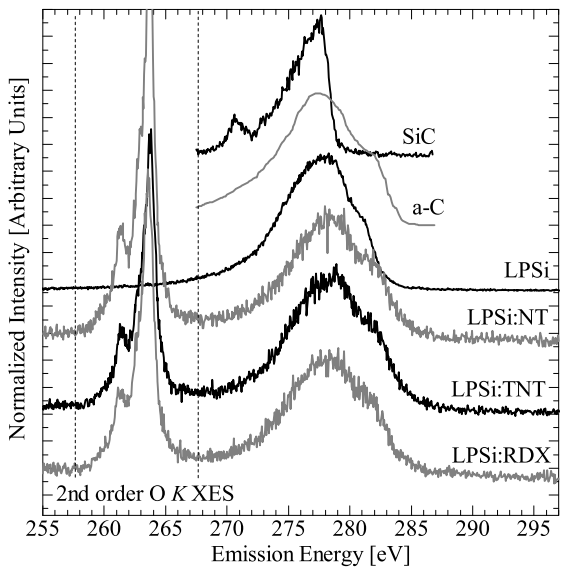}
\caption{\label{fig:carbon} The carbon \textit{K} XES spectra for LPSi treated with various adsorbants compared to amorphous carbon and SiC. The emission features at $\sim$263 eV are due to second-order emission from oxygen.}
\end{figure}

\paragraph{Carbon X-ray Spectroscopy.} We can see from the C \textit{K} XES spectra (see Figure \ref{fig:carbon}) that the interaction between Si and C is fairly weak, indicating that no significant number of Si-C bonds are formed. This is evident from the weak C \textit{K} XES spectra of LPSi:NT, LPSi:TNT and LPSi:RDX. Indeed, there is a strong contribution of the second order O \textit{K} XES in the energy range from 258 to 266 eV, shown in Figure \ref{fig:carbon}. These oxygen features are absent in the spectrum of untreated LPSi and further demonstrate strong oxidation of LPSi upon exposure to nitro-based organic molecules. Further, the actual carbon \textit{K} feature at 270--286 eV in the C \textit{K} XES spectrum of the treated LPSi samples is similar to the spectrum of amorphous carbon (a-C) and very different from the spectrum of SiC. This suggests that the carbon present in the LPSi samples has mostly accumulated from exposure to the ambient environment, and is not strongly bonded to the LPSi surface.

\paragraph{Oxygen X-ray Spectroscopy.} The O \textit{K} XES spectra of LPSi, LPSi:NT, LPSi:TNT, and LPSi:RDX are all quite similar to the spectrum of SiO$_2$, as shown in Figure \ref{fig:oxygen}. The O \textit{1s} XAS bulk sensitive TFY spectra of the LPSi samples are also similar to the rather featureless TFY spectrum of SiO$_2$. The surface sensitive TEY spectra of the treated LPSi samples show some key differences from the TEY spectrum of untreated LPSi and SiO$_2$. In particular, the untreated LPSi lacks the weak pre-edge feature at $\sim$532 eV seen in the spectra of LPSi:NT, LPSi:TNT, and LPSi:RDX. In amino acids, this pre-edge feature is attributed to $\pi^*$ bonding.~\cite{tanaka01} In SiO$_2$ there is a very strong pre-edge feature, although at higher energies ($\sim$ 533 eV) than in treated LPSi. We would like to mention here that the O \textit{1s} XAS TEY spectrum of HPSi is again essentially the same as that from that of HPSi within 1.5 nm of the surface.

The main edge of the O \textit{1s} XAS spectrum of SiO$_2$ (at $\sim$ 540 eV) is also slightly different from the corresponding edge in the LPSi samples. We can use the second derivative method on the O \textit{K} XES and TFY mode O \textit{1s} XAS to estimate the energy gap. The TFY mode O \textit{1s} XAS spectra are more suited to this than the TEY mode O \textit{1s} XAS spectra, since the pre-edge peak in the latter obscure the true onset of the conduction band. This method suggests that the energy gap of SiO$_2$ is 8.7 eV, fairly close to the literature value of 8.9 eV.~\cite{distefano71} We would like to stress that although the O \textit{K} XES can likely probe deeper than the average pore depth, the Si \textit{L}$_{2,3}$ XES of untreated LPSi shows that the bulk material is crystalline silicon. Therefore, all the oxygen present in the untreated LPSi is on the surface, and, therefore, the O \textit{K} XES and \textit{1s} XAS spectra are surface sensitive. We would further like to stress that since Figure \ref{fig:silicon} shows that the silicon in untreated LPSi has electronic structure similar to c-Si, and since Figure \ref{fig:carbon} shows that there is no second-order O \textit{K} XES signal in the C \textit{K} XES spectrum, there is minimal oxidation of the untreated LPSi surface. Since the O \textit{K} XES and \textit{1s} XAS spectra shown in Figure \ref{fig:oxygen} probe the oxygen spectra explicitly, the limited surface oxidation in untreated LPSi is readily apparent. However, an accurate estimate of the relative concentration of oxygen in untreated and treated LPSi cannot be made using these spectra.

To briefly summarize, the X-ray spectroscopy measurements of LPSi show a sensitivity to the species of adsorbed molecule. There is clear evidence that exposure LPSi to NT, TNT, and RDX induces different degrees of oxidization and nitridization of the silicon surface, which causes a shift in the silicon band edge at the surface.

\begin{figure}
\includegraphics[width=2in]{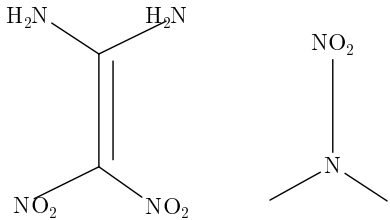}
\caption{\label{fig:fox7} A stylized schematic of 1,1-diamino-2,2-dinitroethylene (DADNE) on the left, and the dimethyl nitroamine (DMNA) molecule on the right.}
\end{figure}

\paragraph{Molecular Dynamics Calculations.} In order to corroborate these experimental observations, we investigated whether dissociative adsorption of molecules containing nitro groups at Si surfaces is plausible. To this end we first simulated the interaction between pure and partially oxidized PSi and volatile molecules with a molecular dynamics approach. This interaction was modeled using Si$_{29}$O$_x$ ($x$=0,...,4) clusters and 1,1-diamino-2,2-dinitroethylene molecule (DADNE) and dimethyl nitroamine (DMNA) molecules shown in Figure \ref{fig:fox7}. These moieties are relatively small to make the \textit{ab initio} calculations feasible, yet, sufficiently complex to assist with adequate interpretation of the experimental data. In particular, the DADNE molecule contains both amino and nitro groups and the DMNA molecule contains both methyl and nitro groups typically found in energetic materials. With respect to the NT, TNT, and RDX studied herein, DADNE has the same single bond C-NO$_2$ structure found in NT and TNT, while DMNA has the same single bond N-NO$_2$ structure present in RDX. Moreover, DADNE yields additional interest as it is a relatively new, highly energetic compound \cite{Bemm_1998} and its mechanical \cite{Zerilli_compression} and electronic \cite{Kuklja_APL_2007,PVS_2007_JCP_FOX7,PVS_2009_Fox7_PRB} properties and mechanisms of decomposition \cite{Politzer_1998,Gindulyte_1999,Kuklja_2007_shear} are actively investigated. 

The Si$_{29}$ cluster has been developed as a model for PSi elsewhere \cite{Gavatrin_Si} and employed previously in, for example, the simulation of atomic force microscopy tip interaction with alkali-halide surfaces.~\cite{PVS_1999_AFM_ASS} The effect of partial oxidation was investigated by calculating the geometrical structures and electronic properties of Si$_{29}$O$_x$ ($x$=1-4) clusters (see Supporting Information for details). In each case the geometrical structure of the Si$_{29}$O$_x$ cluster was fully relaxed using the hybrid B3LYP density functional \cite{Lee_1988,Becke_1993} and Pople's 6-31G(d) basis set, as implemented is the Gaussian 03 package.~\cite{Gaussian_03D02} The energy gain due to reaction with each O atoms is calculated with respect to the half of the total energy of an isolated O$_2$ molecule. We note that although a real PSi surface prepared by HF may have a hydride layer, the coverage of this layer depends on the preparation conditions and the layer itself unstable and easily replaced by other species.~\cite{boukherroub_09} We therefore did not include a surface hydride layer on the Si$_{29}$ cluster as such an addition would greatly increase the degrees of freedom of the system, and consequently the calculation time.

\begin{figure}
\includegraphics[width=3in]{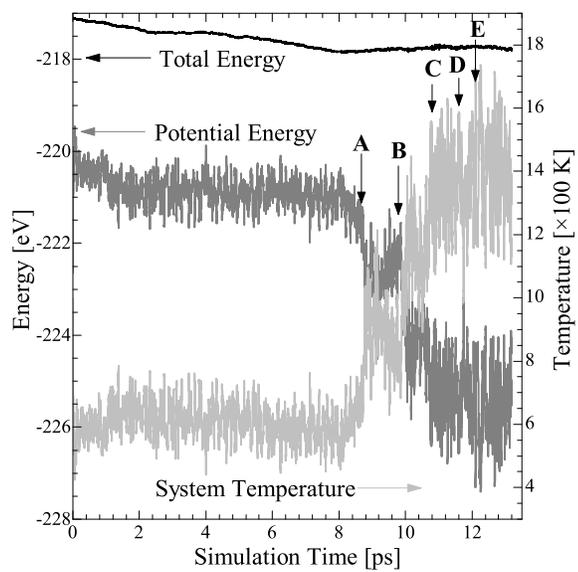}
\caption{\label{fig:calc} Time evolution of the system potential energy (left-most axis) and temperature (right-most axis) for the DADNE molecule interacting with the Si$_{29}$ cluster. The total energy of the system (also left-most axis) is approximately constant throughout. The events ``A'' through ``E'' are described in the text.}
\end{figure}

The time evolution of the system's potential energy and temperature, the latter being equivalent to the system kinetic energy, are shown in Figure \ref{fig:calc}. The two plots anti-correlate, so the total energy of the system remains constant within the numerical noise of these calculations. The Si$_{29}$ cluster and the DADNE molecules are separated at the beginning of the simulation and the interaction between them is negligible up until approximately 8 ps of the simulation time. During this period the cluster is stationary and the molecule, as a whole, moves slowly across the supercell. The initial kinetic energies of atoms were set up so as the system temperature at this stage was approximately twice that of the room temperature (see Figure \ref{fig:calc}).

\begin{figure}
\includegraphics[width=3in]{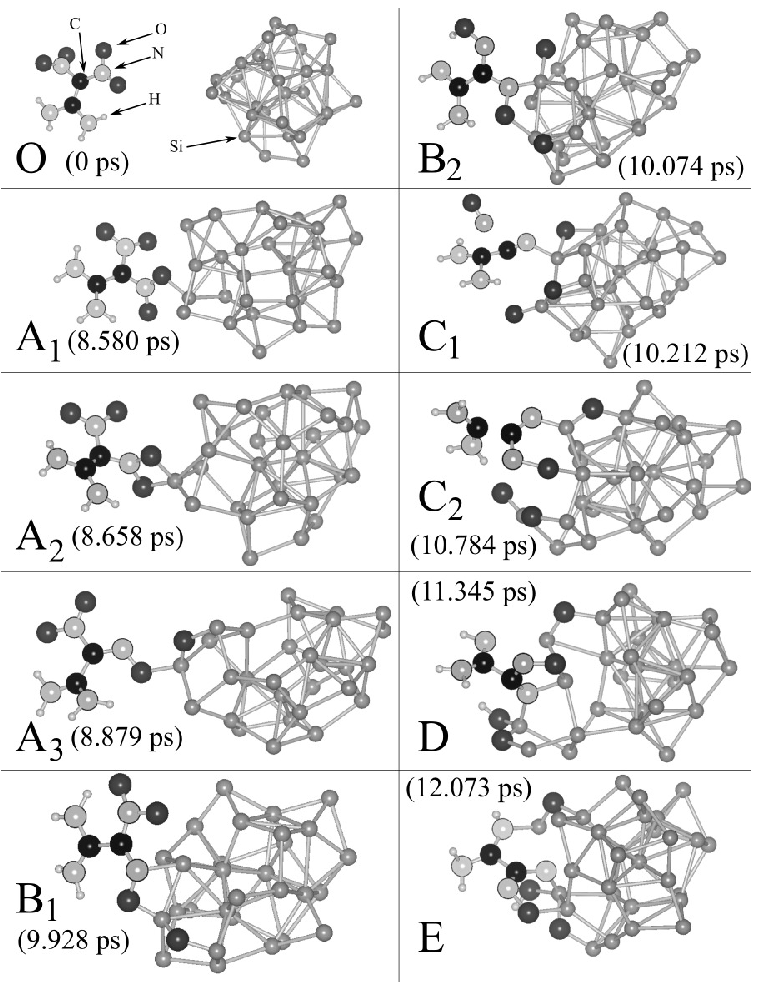}
\caption{\label{fig:dadne} Configurations of the DADNE--Si$_{29}$ system at several key stages during the MD simulation run. Panel \textbf{O} represents the original configuration. The events of subsequent panels are described in the text, the letter of each panel identifies the simulation time step in Figure \ref{fig:calc} when the pictured event occurred.}
\end{figure}

At approximately 8 ps of the simulation time (Figure \ref{fig:calc}), the DADNE molecule approaches the Si$_{29}$ cluster with one of its NO$_2$ groups oriented toward the cluster atoms. The interaction between the cluster and the molecule nitro-groups induces several events indicated as ``A'' in Figure \ref{fig:calc}, which take place consecutively (shown schematically in panels \textbf{A}$_1$ through \textbf{A}$_3$ in Figure \ref{fig:dadne}):
\begin{enumerate}
\item The formation of the first O-Si bond,
\item The formation of the second O-Si bond, whereby a N-O-Si-O cycle is created, and
\item The insertion of one O atom into a neighbouring Si-Si bond, with the formation of a N-O-Si-O-Si fragment. 
\end{enumerate}
During this time the remainder of the DADNE molecule is attached to the Si cluster and its dynamics are coupled to those of the cluster. As the system continues to evolve in time, two events indicated as ``B'' occur (shown schematically in panels \textbf{B}$_1$ and \textbf{B}$_2$ in Figure \ref{fig:dadne}): 
\begin{enumerate}
\item The displacement of a single Si atom towards the DADNE molecule and formation of bonds with an N atom of the first, now decomposed, NO$_2$ group and an O atom of the second, still intact, NO$_2$ group, and
\item The displacement of the same Si atom back towards the cluster so the O-N bond in the NO$_2$ group is broken and new Si-O-Si bonds are formed. 
\end{enumerate}
At this stage, three out of four O atoms of the DADNE molecule are bonded to Si atoms of the cluster and consequent processes involve N as well as C atoms in the remains of the DADNE molecule. In event ``C'', a NO complex, which is the only surviving fragment of the original DADNE nitro groups, splits from the system and becomes a gas-phase NO molecule (shown in panel \textbf{C}$_1$ in Figure \ref{fig:dadne}). Such molecules are often observed among the products of detonated organic explosives. In our case, the NO molecule binds to the surface of the silicon cluster, thus forming a six membered Si-O-N-C-N-Si ring (shown in panel \textbf{C}$_2$ in Figure \ref{fig:dadne}). Events ``D'' and ``E'' correspond to the formation of the first Si-N-Si fragment and the switching of one N-Si bonds, respectively (shown in panels \textbf{D} and \textbf{E} in Figure \ref{fig:dadne}, respectively). Events ``A'' and ``B'' result in substantial potential energy gains, which is transferred to the temperature gains. Potential energy gains of similar magnitude follow events ``C'' and ``E''. However, by this stage the system heats up to over 1400 K (see Figure \ref{fig:calc}), which is less than 300 K below silicon melting temperature. Consequently, further oxidation and nitrogenation events are suppressed.

These results suggest that organic molecules containing nitro groups actively react with amorphous Si,  oxidize and nitridize its surface, and decompose in the process. We note, however, that other molecule structural elements, such as amino-groups and the C=C backbone in the case of DADNE molecules, can remain intact in the process of the Si--molecule interaction. Earlier experimental studies have demonstrated that X-ray spectra of conglomerates of organic molecules are often merely a superposition of the spectra of individual molecules and functional groups.~\cite{boese97} In our case, the dramatic changes observed in the Si \textit{L}$_{2,3}$ XES spectra upon exposure of LPSi to the molecules containing nitro-groups also point to dissociative adsorption of these molecules. Similar dissociative adsorption has been demonstrated theoretically for nitroamine molecules at the Al(111) surface.~\cite{zhou10} 

Similar MD simulations have been carried out for the DMNA molecule and the Si$_{29}$ cluster. In this case, the molecule adsorbs at the surface but, unlike DADNE, the NO$_2$ group does not decompose within the simulation time. This difference in the DADNE-Si$_{29}$ and DMNA-Si$_{29}$ interaction correlates with the difference between the spectral shifts observed for NT and TNT and those observed for RDX. However, due to the limited MD simulation time and a qualitative model of the LPSi used in these calculations, the atomistic origin of the difference in the spectral shifts remain unidentified.

Finally, we considered the effect of partial oxidation on both the electronic structure of the silicon cluster and on the interaction of the DADNE and DMNA molecules with the oxidized silicon cluster Si$_{29}$O$_x$. To keep the calculations feasible, we varied the value of $x$ between 1 and 4 and monitored the changes of the geometrical structures and of the DOS with increasing $x$ (see Supporting Information). The interaction of the DADNE and DMNA molecules with the Si$_{29}$O$_4$ cluster, as modeled using the MD approach, is very similar to that with the Si$_{29}$ cluster: the DADNE molecule gradually decomposes at the  Si$_{29}$O$_4$ cluster surface, while the DMNA molecule remains intact during an equally long simulation time. This is not surprising taking into account that the electronic structure of the Si$_{29}$O$_4$ cluster is very similar to that of pure  Si$_{29}$ cluster in that it contains many non-oxidized surface silicon atoms.

\begin{figure}
\includegraphics[width=3in]{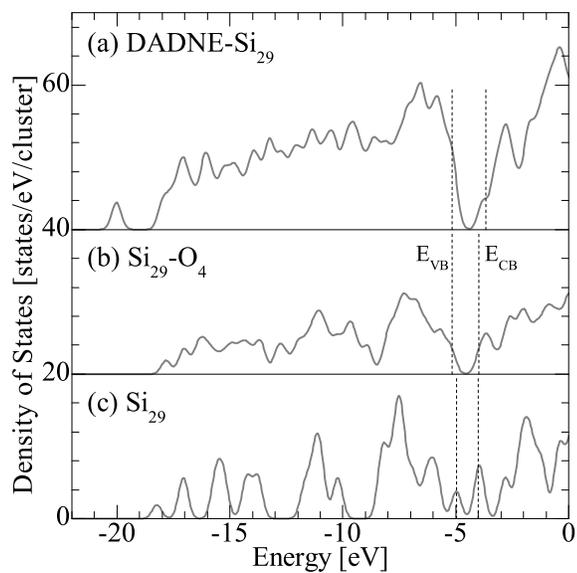}
\caption{\label{fig:si29dos} Calculated total density of states for the fully relaxed geometries of (a) the final state of the dissociative adsorption of a DADNE molecule on the Si$_{29}$ cluster (at a simulation time of 12.073 ps), (b) the Si$_{29}$-O$_4$ cluster, and (c) the Si$_{29}$ cluster. The edges of the valence bands (E$_\mathrm{VB}$) and conduction bands (E$_\mathrm{CB}$) are shown by dotted lines.}
\end{figure}

\paragraph{Electronic Structure Calculations.} In order to analyze the electronic structure changes induced by the decomposition of DADNE on amorphous Si, we have calculated the DOSs for each of the configurations shown in Figure \ref{fig:dadne}, as discussed in the Experimental Section (see Supporting Information). Due to a highly irregular character of the Si cluster surface, detailed analysis of these changes is complicated. Nevertheless, it is apparent that formation of a single Si--O bond (Figure \ref{fig:dadne}, 8.580 ps) results in a noticeable depletion of the density of states near the gap between the highest occupied and lowest unoccupied states (referred to as HOMO-LUMO gap). The gap shifts to the lower energies when two oxygen atoms bind to the same Si atom (Figure \ref{fig:dadne}, 8.658 ps) and widens when one of these oxygens also binds to another Si atoms (Figure \ref{fig:dadne}, 8.879 ps). As DADNE decomposition progresses, the lower valence band states become apparent, the upper valence band gradually becomes featureless (similar to that of amorphous SiO$_2$), and the HOMO-LUMO gap increases, as shown in Figure \ref{fig:si29dos}.  

The changes in electronic structure due to the DADNE decomposition on Si$_{29}$-O$_4$ are similar to those arising due to the DADNE decomposition on Si$_{29}$, as illustrated by the DOS calculated for the Si$_{29}$O$_4$ and Si$_{29}$-DADNE systems (Figure \ref{fig:si29dos}). These changes support the experimental findings (see Figure \ref{fig:silicongap}) that treatment with various molecules shifts the Si \textit{L}$_{2,3}$ XES in energy but does not significantly change the shape of the emission. Importantly, the valence band edge is shifted to lower energies with respect to pristine Si$_{29}$ in both the Si$_{29}$-DADNE and Si$_{29}$O$_x$ systems, while only the Si$_{29}$-DADNE system shows a shift to the higher energies in the conduction band edge as compared to the pristine Si$_{29}$ cluster. This supports the findings that treatment with various molecules can cause a greater change in the surface electronic structure than O$_2$-induced oxidation alone.

Interestingly, the overall energy gain due dissociative DADNE decomposition (see Supporting Information) is $\sim$12 eV, which is comparable to the overall energy gain upon reaction of Si$_{29}$ with two oxygen molecules and formation of the Si$_{29}$O$_4$ cluster ($\sim$12.8 eV). However, the former, according to our simulations, takes place rapidly and spontaneously at temperatures much lower than those used for Si oxidation. Hence, one can expect that reaction of the LPSi with the molecules containing nitro groups proceeds more rapidly than reaction with molecular O$_2$.

\paragraph{Discussion.} The observed effects of the adsorption of nitro-based molecules on the LPSi substrate are summarized in Table \ref{tbl:ratios}. What is the mechanism behind these effects? Si-Si bonds in PSi can be broken to form Si-C bonds,~\cite{song_98} Si-N bonds tend to replace Si-C bonds,~\cite{chen05} and Si-O bonds are even more energetically favourable.~\cite{scopel02} Since the observed spectral shifts are due not only to the functional groups of the adsorbed molecule (i.e. the differences between the spectra of LPSi:RDX and LPSi:TNT) but also the affinity of that molecule to be absorbed (i.e. the differences between the spectra of LPSi:TNT and LPSi:NT), we can attempt to clarify the mechanisms behind our observations.

\begin{table}
\begin{tabular}{c|c|c|c|c}
 & E$_g$ [$\pm$ 0.5 eV] & $\Delta$E$_\mathrm{vb}$ [$\pm$ 0.1 eV] & $\Delta$E$_\mathrm{cb}$ [$\pm$ 0.1 eV] & C : O [$\pm$ 0.01]\\
\hline
LPSi:NT & 8.6 & 0.0 & 0.3 & 1.31 \\
LPSi:TNT & 7.2 & 1.5 & 0.4 & 1.75 \\
LPSi:RDX & 9.9 & -0.8 & 0.4 & 1.48 \\
\end{tabular}
\caption{\label{tbl:ratios} Summary of the differences in X-ray spectra. E$_g$ is the estimated energy gap, derived from the deviations in the Si \textit{L}$_{2,3}$ XES and Si \textit{2p} XAS from those of SiO$_2$ (with a measured energy gap of 8.7 eV).  $\Delta$E$_\mathrm{vb}$ and $\Delta$E$_\mathrm{cb}$ are the shifts in the valence and conduction band edges, respectively. C : O is the estimated ratio between the quantity of surface carbon and oxygen, errors in this quantity were estimated based on the amount of statistical noise in the measurement and the inherent energy uncertainty.}
\end{table}

With this in mind, the C : O ratio in Table \ref{tbl:ratios} (the ratio between the total intensity of the C \textit{K} XES and the second order O \textit{K} XES signal visible in the same spectrum) can be used as an estimate of the relative amount of carbon compared to oxygen between the treated LPSi samples. From the C \textit{K} XES spectra, this ratio is the lowest for LPSi:NT, and the highest for LPSi:TNT, with LPSi:RDX somewhere in between. The C : O ratio per molecule (see Figure \ref{fig:molecules}) is $\dfrac{7}{2}$ for NT, $\dfrac{7}{6}$ for TNT, and $\dfrac{3}{6}$ for RDX. However the actual adsorbed C : O ratio for the treated LPSi depends on the adsorption affinity for each molecule.

Our MD simulations show that C attaches to the Si surface via N (in DADNE) or NO$_2$ (in DMNA) groups. Therefore, the larger the number of NO$_2$ groups per cyclic ring, the greater the chance that the ring will stay attached to the Si surface. While a single N/NO$_2$ - Si bond can leave the aromatic ring out of contact with the Si surface (in the case of NT), two or more N/NO$_2$ - Si bonds will bring the aromatic ring at least partially in plane with the Si surface (in the case of TNT). This close contact with the Si surface presents the opportunity for more bonds (Si - N or even Si - C) to form. Therefore even though the product of the molecular C : O ratio and the number of NO$_2$ groups is the same for both NT and TNT ($\dfrac{7}{2} \times 1$ for NT, and $\dfrac{7}{6} \times 3$ for TNT) we expect more TNT to be retained on the Si surface than NT. In fact, our experimental data suggests that $\sim$1.33 TNT molecules are adsorbed for every NT molecule (see Table \ref{tbl:ratios}).

Since RDX and TNT have the same number of NO$_2$ groups, all else being equal, based on the product of the molecular C : O ratios and the number of NO$_2$ groups we would expect that LPSi would adsorb $\dfrac{7}{3}$ TNT molecules for every RDX molecule (the ratio of $\dfrac{7}{6} \times 3$ for TNT and $\dfrac{3}{6} \times 3$ for RDX). However, the cyclic part of the RDX molecules can, in principle, adsorb to the Si surface via either NO$_2$ or, if the NO$_2$ groups dissociate, via the N atoms in the cyclic ring. Hence the number of potential bonding sites in RDX is twice that of TNT. This implies the ratio of adsorbed TNT molecules to adsorbed RDX molecules should be closer to $\dfrac{7}{6}$. This is quite close  to the ratio of the measured adsorbed C : O for TNT and RDX ($\sim$ 1.18) in Table \ref{tbl:ratios}.

\section{Conclusions}

We have investigated the interaction of low porosity Si with saturated vapours of NT, TNT, and RDX molecules, all of which contain nitro groups. The changes in the surface electronic structure have been characterized using a combination of spectroscopic measurements for both pre-oxidized LPSi and LPSi exposed to the NT, TNT, or RDX vapours (see Figure \ref{fig:molecules}). The Si \textit{L}$_{2,3}$ XES and \textit{2p} XAS, the C \textit{K} XES, and the O \textit{K} XES and \textit{1s} XAS spectra show changes in the surface electronic structure of LPSi, which points to dissociative adsorption of the molecules. Our \textit{ab initio} molecular dynamics calculations support the feasibility of this dissociative adsorption and provide insight into mechanisms on the atomic scale of silicon oxidation and nitrogenation. These calculations also suggest pathways of simultaneous decomposition of the adsorbed molecules. Furthermore, our \textit{ab initio} electronic structure calculations support the feasibility of dissociative adsorption driving changes in the valence and conduction band edges.

Importantly, we find that, in spite of the similarity in the dissociative nature of the molecular adsorption, the changes to the LPSi electronic structure are clearly molecule-specific. In particular, the estimated values of the surface energy gap decreases as the number of nitro functional groups increases for adsorbants containing aromatic rings, such as NT and TNT. On the contrary, the surface energy gap increases for adsorbants containing nitroamine groups, such as RDX. This suggests the relative energy shifts and, therefore, selectivity towards the adsorbed molecules are not due to adsorption of nitro groups alone, but due to adsorption of nitroamine groups (from the dissociated RDX molecules), and aromatic rings or methyl groups (from the dissociated NT or TNT molecules) on the surface of LPSi. 

The strong changes in the electronic structure of LPSi that are observed after adsorption of a particular molecule species support the feasibility of LPSi as a material for the detection and identification of trace amounts of airborne explosive molecules.

\section{Experimental Section}

\paragraph{Preparation of PSi.} The PSi used in this experiment was prepared by the common etching technique, described in detail elsewhere.~\cite{levitsky07,cullis97} In brief, we fabricated PSi by electrochemical etching of crystalline silicon  (c-Si) with a hydrofluoric acid (HF) solution. High porosity Si (HPSi) and low porosity Si (LPSi) with surface pore densities of 75.2\% and 43.5\%, respectively, were prepared. The PSi surfaces were oxidized at 900$^\circ$ C for 20 minutes and then exposed to saturated vapours of TNT, NT, or RDX. For more details see the supplemental information.

\paragraph{X-ray Spectroscopy.} We performed X-ray emission spectroscopy (XES) and X-ray absorption spectroscopy (XAS), probing the occupied and unoccupied states, respectively. The XAS measurements were performed in both the bulk-sensitive total fluorescence yield (TFY) mode, and the surface-sensitive total electron yield (TEY) mode. The Si \textit{2p} XAS and C, O \textit{1s} XAS measurements were performed at the Spherical Grating Monochromator \cite{regier07} and Variable Line-spacing Planar Grating Monochromator \cite{hu07} beamlines at the Canadian Light Source, respectively. All XES measurements were performed at beamline 8.0.1 \cite{jia95} at the Advanced Light Source. The samples were removed from the saturated vapour and measured under ultra-high vacuum with no further preparation.

\paragraph{Theoretical Calculations.} The DADNE--Si$_{29}$O$_x$ and DMNA--Si$_{29}$O$_x$ ($x$=0,4) systems were investigated using the \textit{ab initio} molecular dynamics (MD) method, in which atoms obey Newtonian equations of motion. The system's total energies and the forces acting on atoms are calculated using the generalized gradient corrected density functional by Perdew, Burke, and Ernzerhof (PBE) \cite{PBE_1996_PRL} and the projected augmented waves method.~\cite{Blochl_1994_PRB_PAW} The calculations are performed using the Vienna \textit{ab initio} simulation package (VASP).~\cite{Kresse_1996_PRB_VASP} The system was modeled using periodic boundary conditions and a cubic supercell with the lattice constant of 18 \AA. A single $k$-point ($\Gamma$) has been used for Brillouin zone integration. The constant total energy ensemble and 1 fs timestep were used for the MD production runs. 

The electronic structure modifications of these systems have been analyzed using the densities of states (DOS) calculated for judiciously selected MD snapshot configurations. In other to eliminate the effects of the thermal noise, the energy of each configuration has been fully minimized with respect to the atomic coordinates at the B3LYP/6-31G(d) level of theory and using the Gaussian 03 package.~\cite{Gaussian_03D02} The DOS is obtained as a convolution of Gaussian-type functions (FWHM=0.5 eV), each representing a single one-electron state.

\paragraph{Acknowledgements.}We acknowledge support of the Natural Sciences and Engineering Research Council of Canada (NSERC), the Canada Research Chair program, and the Russian Science Foundation for Basic Research Project No. 11-02-00022. Some of the research described in this paper was performed at the Canadian Light Source, which is supported by the NSERC, the National Research Council (NSC) Canada, the Canadian Institutes of Health Research (CIHR), the Province of Saskatchewan, Western Economic Diversification Canada, and the University of Saskatchewan. PVS is supported by the Royal Society. PVS thanks M. M. Kuklja for stimulating discussions. Calculations have been performed at the HECTOR facility (access provided via the Materials Chemistry Consortium, EPSRC grant EP/F067496) and EMSL, a national scientific user facility sponsored by the Department of Energy's Office of Biological and Environmental Research and located at Pacific Northwest National Laboratory. The Advanced Light Source is supported by the Director, Office of Science, Office of Basic Energy Sciences, of the U.S. Department of Energy under Contract No. DE-AC02-05CH11231.

\bibliographystyle{unsrt}
\bibliography{levitsky}

\end{document}